\documentclass[12pt]{article}

\usepackage{amsmath}
\allowdisplaybreaks

\usepackage{amssymb}
\usepackage{bm}

\begin{document}

\baselineskip=17.0pt plus 0.2pt minus 0.1pt

%%%%%%%%%% Private macros %%%%%%%%%%%%
\makeatletter
\@addtoreset{equation}{section}
\makeatother

\renewcommand{\theequation}{\thesection.\arabic{equation}}
\renewcommand{\thefootnote}{\fnsymbol{footnote}}
\newcommand{\calA}{{\cal A}}
\newcommand{\calB}{{\cal B}}
\newcommand{\calE}{{\cal E}}
\newcommand{\calP}{{\cal P}}
\newcommand{\calM}{{\cal M}}
\newcommand{\calN}{{\cal N}}
\newcommand{\calV}{{\cal V}}
\newcommand{\calK}{{\cal K}}
\newcommand{\calF}{{\cal F}}
\newcommand{\calH}{{\cal H}}
\newcommand{\calT}{{\cal T}}
\newcommand{\calU}{{\cal U}}
\newcommand{\calY}{{\cal Y}}
\newcommand{\calW}{{\cal W}}
\newcommand{\calL}{{\cal L}}
\newcommand{\calD}{{\cal D}}
\newcommand{\calO}{{\cal O}}
\newcommand{\calQ}{{\cal Q}}
\newcommand{\calS}{{\cal S}}
\newcommand{\QB}{{\cal Q}_\text{B}}
\newcommand{\nn}{\nonumber}
\newcommand{\deriv}[2]{\frac{d #1}{d#2}}
\newcommand{\eps}{\varepsilon}
\newcommand{\ds}{\displaystyle}
\newcommand{\Fe}{F_{\eps}}
\newcommand{\Ke}{K_{\eps}}
\newcommand{\Psie}{\Psi_{\eps}}
\newcommand{\CR}[2]{\left[#1,#2\right]}
%%%%%%%%%%%%%%%%%%%%%%%%%%%%%%%%%
\newcommand{\tr}{\mathop{\rm tr}}
\newcommand{\Tr}{\mathop{\rm Tr}}
\newcommand{\wN}{N}
\newcommand{\swN}{\calN}
\newcommand{\p}{\partial}
\newcommand{\wh}[1]{\widehat{#1}}
\newcommand{\abs}[1]{\left| #1\right|}
\newcommand{\KeR}{$\Ke$-regularization}
\newcommand{\Ker}{$\Ke$-reg.}
\newcommand{\Utv}{U_\text{tv}}
\newcommand{\wt}[1]{\widetilde{#1}}
\newcommand{\KBc}{$KB\,c$}
\newcommand{\VEV}[1]{\left\langle #1\right\rangle}
\newcommand{\Drv}[2]{\frac{\p #1}{\p #2}}
%%%%%%%%%%%%%%%%%%%%%%%%%%%%%%%%%%%%%%

\begin{titlepage}

\title{
\hfill\parbox{3cm}{\normalsize KUNS-2371}\\[1cm]
{\Large\bf
Winding Number in String Field Theory
}}

\author{
Hiroyuki {\sc Hata}\footnote{
{\tt hata@gauge.scphys.kyoto-u.ac.jp}}
\ and
Toshiko {\sc Kojita}\footnote{
{\tt kojita@gauge.scphys.kyoto-u.ac.jp}}
\\[7mm]
{\it
Department of Physics, Kyoto University, Kyoto 606-8502, Japan
}
}

\date{{\normalsize November 2011}}
\maketitle

\begin{abstract}
\normalsize
Motivated by the similarity between cubic string field theory (CSFT)
and the Chern-Simons theory in three dimensions, we study the
possibility of interpreting $\swN=(\pi^2/3)\int\!(U\QB U^{-1})^3$ as a
kind of winding number in CSFT taking quantized values.
In particular, we focus on the expression of $\swN$ as the integration
of a BRST-exact quantity, $\swN=\int\!\QB\calA$, which vanishes
identically in naive treatments.
For realizing non-trivial $\swN$, we need a regularization for
divergences from the zero eigenvalue of the operator $K$ in the
\KBc\ algebra. This regularization must at same time violate the
BRST-exactness of the integrand of $\swN$.
By adopting the regularization of shifting $K$ by a positive
infinitesimal, we obtain the desired value
$\swN[(\Utv)^{\pm 1}]=\mp 1$ for $\Utv$ corresponding to the tachyon
vacuum. However, we find that $\swN[(\Utv)^{\pm 2}]$ differs from
$\mp 2$, the value expected from the additive law of $\swN$.
This result may be understood from the fact that $\Psi=U\QB U^{-1}$
with $U=(\Utv)^{\pm 2}$ does not satisfy the CSFT EOM in the strong
sense and hence is not truly a pure-gauge in our regularization.

\end{abstract}

\thispagestyle{empty}
\end{titlepage}

\section{Introduction}
\label{sec:intro}

Cubic open string field theory (CSFT) \cite{CSFT} has strong
resemblances in its algebraic structure with the Chern-Simons (CS)
theory.
In fact, the action and the gauge transformation of CSFT,
\begin{align}
S&=\frac{1}{g_o^2}\int\left(
\frac12\Psi*\QB\Psi+\frac13\Psi*\Psi*\Psi\right) ,
\label{eq:S_CSFT}
\\
\delta\Psi&=\QB\Lambda+\Psi*\Lambda-\Lambda*\Psi ,
\label{eq:dPsi}
\end{align}
are obtained from the action of the CS theory,
\begin{equation}
S_\text{CS}=\frac{k}{2\pi}\int_M\tr\left(
\frac12 A\wedge dA+\frac13 A\wedge A\wedge A\right) ,
\label{eq:S_CS}
\end{equation}
and its gauge transformation by the following replacements:
\begin{equation}
A\to \Psi,\quad
d\to\QB,\quad
\wedge\to *,\quad
\int_M\tr \to \int .
\end{equation}
The invariance of the CSFT action \eqref{eq:S_CSFT} under the
infinitesimal gauge transformation \eqref{eq:dPsi} is due to that the
three basic operations $\QB$, $*$ and $\int$ in CSFT enjoy the same
algebraic properties as those of $d$, $\wedge$ and $\int_M\tr$ in
the CS theory:
\begin{align}
\QB^2&=0 ,
\\
\QB\left(\Phi*\Sigma\right)&=\left(\QB\Phi\right)*\Sigma
+(-1)^{g(\Phi)}\Phi*\left(\QB\Sigma\right) ,
\\
\left(\Phi*\Sigma\right)*\Xi&= \Phi *\left(\Sigma*\Xi\right) ,
\\
\int \Phi*\Sigma&=(-1)^{g(\Phi)g(\Sigma)}\int\Sigma*\Phi ,
\label{eq:cyclicity}
\\
\int\!\QB\Phi&=0 ,
\label{eq:intQBPsi=0}
\end{align}
where $\Phi$, $\Sigma$ and $\Xi$ are arbitrary string fields and
$g(\Phi)$ is the ghost number of $\Phi$.

Under the finite gauge transformation,
$A\to g\left(d+A\right)g^{-1}$, by a gauge group valued function
$g(x)$, the CS action \eqref{eq:S_CS} is transformed as
\begin{equation}
S_\text{CS}\to S_\text{CS}-2\pi k \wN[g] ,
\label{eq:SCStoSCS-N}
\end{equation}
where $\wN[g]$ is the winding number of the mapping $g(x)$ from the
manifold $M$ to the gauge group:
\begin{equation}
\wN[g]=\frac{1}{24\pi^2}\int_M\tr\left(gdg^{-1}\right)^3 .
\label{eq:wN}
\end{equation}
Due to this property, the coefficient $k$ multiplying the CS action
\eqref{eq:S_CS} is required to be an integer (the level of the
theory).

The CSFT has quite the same property under a finite gauge
transformation:
\begin{equation}
\Psi\to U\left(\QB+\Psi\right)U^{-1} ,
\label{eq:Psi^U}
\end{equation}
where all the products should be regarded as the star product $*$, and
$U$ is given by
$U=e^{-\Lambda}=1-\Lambda+(1/2)\Lambda^2-\cdots$ with $1$ being
the identity string field.
Under \eqref{eq:Psi^U}, the CSFT action is transformed as\footnote{
On the RHS of \eqref{eq:StoS-N}, we have omitted the term
$-\frac12\int\!\QB\!\left[(\QB U^{-1})U\Psi\right]$ (this is the case
also in \eqref{eq:SCStoSCS-N} for the CS theory).
This term is equal to zero if we can use the property
\eqref{eq:intQBPsi=0}, and, for a pure-gauge $\Psi$, it is nothing but
the extra term $\Delta\swN$ \eqref{eq:DN} in the additive law of
$\swN$.}
\begin{equation}
S\to S -\frac{1}{2\pi^2 g_o^2}\,\swN[U] ,
\label{eq:StoS-N}
\end{equation}
with $\swN[U]$ given by
\begin{equation}
\swN[U]=\frac{\pi^2}{3}\int\left(U\QB U^{-1}\right)^3 .
\label{eq:swN}
\end{equation}
Recently, various translationally invariant exact solutions in CSFT
have been constructed in the pure-gauge form $\Psi=U\QB U^{-1}$
\cite{TakaTani,Schnabl,Okawa,ES,MS}.
For such solutions, their energy density is given by
$\swN/(2\pi^2g_o^2)$.\footnote{
We put the space-time volume equal to one in this paper.
The factor $\pi^2/3$ multiplying \eqref{eq:swN} has been chosen so
that $\swN=-1$ for the tachyon vacuum solution.}

Then, several questions naturally arise.
The first question would be whether $\swN$ \eqref{eq:swN} has an
interpretation as the ``winding number'' taking quantized values.
If so, windings in what sense does $\swN$ count?
Certainly, $\swN$ is a topological quantity invariant under a
small deformation of $U$ similarly to $\wN$ \eqref{eq:wN}.
If $\swN$ takes integer values in some unit, should the inverse of the
open string coupling constant $1/g_o^2$ be quantized as in the case of
the CS theory?

The purpose of this paper is to study whether we can really interpret
$\swN$ as a kind of winding number taking quantized values.
In particular, we focus on the following expression of $\swN$:
\begin{equation}
\swN=\int\!\QB\calA ,
\label{eq:swN=intQBcalA}
\end{equation}
where the quantity $\calA$ is given explicitly in \eqref{eq:calA=}
One might think that the RHS of \eqref{eq:swN=intQBcalA} is equal to
zero since the integration of a BRST-exact quantity is usually
regarded to vanish as given in \eqref{eq:intQBPsi=0}.
However, eq.\ \eqref{eq:intQBPsi=0} is not an axiom of CSFT but is an
equation to be proved.
We already know that $\swN$ is non-vanishing for $U$ corresponding to
the tachyon vacuum solution $\Psi=U\QB U^{-1}$, and hence the RHS must
also be so.
In fact, we will see in Sec.\ \ref{sec:calBinKeR} that the RHS of
\eqref{eq:swN=intQBcalA} can be non-vanishing due to
singularities existent in $\calA$.
We add that the formula \eqref{eq:swN=intQBcalA} is practically useful
for calculating $\swN$ for various $U$'s.
Finally, eq.\ \eqref{eq:swN=intQBcalA} suggests us to rewrite its RHS
further as an integration of $\calA$ on the ``boundary'':
$\int_\calM\QB\calA=\int_{\p\calM}\calA$.
It is an interesting problem (but is beyond the scope of this paper)
to clarify whether such a formula exists, and if so, what the
``manifold'' $\calM$ and its ``boundary'' $\p\calM$ are.
This is important for understanding the topological meaning of $\swN$
in CSFT.

Let us mention here the correspondent of \eqref{eq:swN=intQBcalA},
$N=\int_M dG$, in the CS side.
The two-form $G$ is given in terms of the Lie algebra valued function
$\phi(x)$ in $g(x)=e^{i\phi(x)}$, and $\phi(x)$ has, in general,
singularities in $M$. We can evaluate $N$ as $N=\int_{\p M}G$ with $\p
M$ being the singular points of $\phi(x)$.
Let us take the simplest example of the $SU(2)$ gauge group and the
hedgehog type $g(x)$ on $M=S^3$;
$g(x)=\exp\left(if(r)\,\wh{\bm{x}}\cdot\bm{\tau}\right)$ with
$r=\abs{\bm{x}}$ and $\wh{\bm{x}}=\bm{x}/r$.
The regularity of $g(x)$ at the origin and the infinity requests that
$f(0)$ and $f(\infty)$ be integer multiples of $\pi$. For this $g(x)$,
we have
\begin{equation}
N=\int\!d^3x\,\bm{\nabla}\!\cdot\!\left[
\frac{2f-\sin 2f}{8\pi^2 r^2}\,\wh{\bm{x}}\right]
=\frac{1}{\pi}\left(f(\infty)-f(0)\right) .
\end{equation}

In the rest of this section, let us explain in some detail the results
of our analysis. As we mentioned below \eqref{eq:swN}, exact classical
solutions, including the tachyon vacuum solution, have been
constructed in the pure-gauge form $\Psi=U\QB U^{-1}$.
Their construction is most concisely given in terms of the \KBc\
algebra \cite{Okawa,ES}, which we summarize in Appendix
\ref{app:KBc}.
The points concerning \eqref{eq:swN=intQBcalA} is that there appears
$1/K$ on its RHS and that the eigenvalue of the operator $K$ is
non-negative and, in particular, $K$ has a zero eigenvalue.\footnote{
Though there seems to be no rigorous proof for this fact, concrete
calculations of various correlators support it.}
Furthermore, the existence of the zero eigenvalue of $K$ endangers the
simple identity $K(1/K)=1$ and hence the validity of algebraic
manipulations in the \KBc\ algebra if we use the Schwinger
parametrization for $1/K$:
\begin{equation}
K\frac{1}{K}=K\int_0^\infty\!dt\,e^{-tK}
=1-e^{-\infty\,K} .
\label{eq:K(1/K)}
\end{equation}
It is a subtle problem whether the last term can be dropped if $K$ has
a zero eigenvalue. Therefore, some kind of regularization for the
zero eigenvalue needs to be  introduced for calculating (or defining)
\eqref{eq:swN=intQBcalA} in a well-defined manner.
In this paper, we adopt the regularization of shifting $K$ by an
infinitesimal positive constant $\eps$:
\begin{equation}
K \to \Ke=K+\eps .
\label{eq:KtoKe=K+eps}
\end{equation}
Namely, we lift the eigenvalues of $K$ by $\eps$. In the end of the
calculation, we take the limit $\eps\to +0$.
We call this regularization ``\KeR'' hereafter.
There may be other kinds of regularizations for
\eqref{eq:swN=intQBcalA}. However, introducing the upper cutoff to the
integration of the Schwinger parameter for $1/K$ does not help making
\eqref{eq:swN=intQBcalA} non-vanishing.

If we apply the \KeR\ to \eqref{eq:swN=intQBcalA}, BRST-exactness
of the integrand is violated by $O(\eps)$.\footnote{
In our \KeR, we first evaluate the operations of $\QB$, and then
replace all $K$ in the resultant expression with $\Ke$.
}
We find that this does in fact leads to a non-vanishing and the
expected value of $\swN$ for the tachyon vacuum solution.
More concretely, we have $\swN=\int(\QB\calA)_{K\to\Ke}=\eps\int\calW
=\eps\times O(1/\eps)=-1$, where $\calW$
depends on $\eps$ through $\Ke$.

We also examine $\swN$ \eqref{eq:swN=intQBcalA} for other
$U$'s than that of the tachyon vacuum for testing whether
$\swN$ is really quantized. For this purpose we note the relation
\begin{equation}
\swN[UV]=\swN[U]+\swN[V]+\int\!\QB(\cdots) ,
\label{eq:N=N1+N2+intQB}
\end{equation}
where $(\cdots)$ in the last term is given in terms of $U$ and
$V$ (see \eqref{eq:DN} for a precise expression).
This equation implies the additive property of $\swN$,
if we can discard the last term given as the integration of a
BRST-exact quantity. Therefore, for $(\Utv)^n$
($n=\pm 1,\pm2,\ldots$) with $\Utv$ describing the tachyon vacuum, we
would have $\swN[(\Utv)^n]=n\,\swN[\Utv]=-n$.
We calculate $\swN[(\Utv)^n]$ for $n=\pm 1$ and $\pm 2$ in the \KeR\
to find that
\begin{equation}
\swN\bigl[(\Utv)^n\bigr]=
\begin{cases}
2-2\pi^2 & (n=-2)\\
1 & (n=-1)\\
-1 & (n=1; \text{tachyon vacuum})\\
-2+2\pi^2 &(n=2)
\end{cases} .
\label{eq:swN[U_n]=}
\end{equation}
Namely, we get the expected result for $n=-1$, but the results for
$n=\pm 2$ are anomalous and signal the violation of the additive
property of $\swN$.
We calculate the last term of \eqref{eq:N=N1+N2+intQB} in the \KeR\
and find that it is non-vanishing and accounts for the anomalous part
of \eqref{eq:swN[U_n]=} for $n=\pm 2$.
Does our result \eqref{eq:swN[U_n]=} give a counterexample to our
expectation that $\swN$ is quantized?
Our answer would be no.
We examine whether the classical solution
$\Psi_n=(\Utv)^n\QB(\Utv)^{-n}\bigr|_{K\to\Ke}$ in the \KeR\ satisfies
the EOM in the strong sense, namely, whether
$\int\!\Psi_n*\left(\QB\Psi_n+\Psi_n *\Psi_n\right)=0$ holds.
In other words, we test whether $\Psi_n$ is really a pure-gauge.
We find that the EOM in the strong sense holds for $n=\pm 1$,
but it is violated for $n=\pm 2$.
Therefore, our result \eqref{eq:swN[U_n]=} for $n=\pm 2$ cannot be
regarded as a counterexample.

The organization of the rest of this paper is as follows.
In Sec.\ 2, we derive \eqref{eq:swN=intQBcalA} for a generic $U$, and
its concrete expression in the \KBc\ algebra.
In Sec.\ 3, we calculate $\swN$ for $\Utv$ of the tachyon vacuum in
the \KeR. Then, in Sec.\ 4, we present our analysis on $\swN$, its
additive law and the EOM for $(\Utv)^n$. The final section (Sec.\ 5)
is devoted to a summary and discussions on future problems.
In Appendix \ref{app:KBc}, we summarize the \KBc\ algebra and the
correlators used in the text. In Appendix B, we present another way of
calculating \eqref{eq:swN=intQBcalA} for the tachyon vacuum.

\section{$\bm{\swN}$ as the integration of a BRST-exact quantity}

In this section, we first show \eqref{eq:swN=intQBcalA}, namely, that
$\swN$ \eqref{eq:swN} is given as the integration of a BRST-exact
quantity $\QB\calA$. Then, we derive the expression of $\QB\calA$ for
$U=1-F(K)BcF(K)$, which has been used for constructions
of classical solutions in the \KBc\ algebra.
In the rest this paper, we put the open string coupling constant
$g_o$ equal to one. We often omit $*$ for the string field product
unless confusion occurs.

\subsection{Derivation of $\bm{\swN=\int\!\QB\calA}$}

Let us consider $\Psi$ given in a pure-gauge form $\Psi=U\QB
U^{-1}$. For this $\Psi$, we introduce $\Psi_s=U_s\QB U_s^{-1}$ with
$U_s$ carrying a parameter $s$ ($0\le s\le 1$) and interpolating $U$
and $1$; $U_{s\,=\,1}=U$ and $U_{s\,=\,0}=1$.
Then, we can show that $\calA$ given by
\begin{equation}
\calA=\pi^2\int_0^1\!ds\,\Psi_s *\deriv{\Psi_s}{s} ,
\label{eq:calA=}
\end{equation}
satisfies \eqref{eq:swN=intQBcalA}. The proof goes as follows:
\begin{align}
\frac{1}{\pi^2}\swN&=\frac13\int\Psi^3
=\frac13\int_0^1\!ds\deriv{}{s}\int\!\Psi_s^3
=\int_0^1\!ds\int\Psi_s*\deriv{\Psi_s}{s}*\Psi_s
\nn\\
&=\int_0^1\!ds\int\!\left(\Psi_s*\deriv{}{s}(\Psi_s)^2
-(\Psi_s)^2*\deriv{}{s}\Psi_s\right)
\nn\\
&=-\int_0^1\!ds\int\!\left(\Psi_s*\deriv{}{s}\QB\Psi_s
-(\QB\Psi_s)*\deriv{}{s}\Psi_s\right)
=\int_0^1\!ds\int\QB\!\left(\Psi_s *\deriv{\Psi_s}{s}\right),
\label{eq:proof}
\end{align}
where we have used the cyclicity \eqref{eq:cyclicity} and the EOM
satisfied by $\Psi_s$: $\QB\Psi_s+(\Psi_s)^2=0$.\footnote{
This proof remains valid for $\Psi$ not restricted to the pure-gauge
form if we can construct $\Psi_s$ satisfying the EOM for all $s$.
}

\subsection{$\bm{\swN}$ in the $\bm{KBc}$ algebra}

We wish to calculate $\swN$ given in the form \eqref{eq:swN=intQBcalA}
with $\calA$ given by \eqref{eq:calA=} for various $U$, in particular,
for $U$ corresponding to the tachyon vacuum.
This calculation will be carried out in Secs.\ 3 and 4.
Here, as a preparation, we present a convenient expression of the RHS
of \eqref{eq:swN=intQBcalA}. This is obtained by interchanging the
order of the $s$ and the CSFT integrations:
\begin{equation}
\swN=\pi^2\int_0^1\!ds\,\calB(s) ,
\label{eq:swN=intcalB}
\end{equation}
with $\calB(s)$ given again as the integration of a BRST-exact quantity:
\begin{equation}
\calB(s)=\int\!\QB\!\left(\Psi_s*\deriv{\Psi_s}{s}\right) .
\label{eq:calB=intQB()}
\end{equation}
By following the manipulation of \eqref{eq:proof} in the reverse way,
we obtain another expression of $\calB(s)$:
\begin{equation}
\calB(s)=\int\!\deriv{\Psi_s}{s}*\Psi_s*\Psi_s .
\label{eq:calB=another}
\end{equation}
It should be noted that the integrand of \eqref{eq:calB=another} is,
though not manifest, BRST-exact.

Next, we present concrete expressions of $\calB(s)$ for $U$
which has been adopted in the construction of classical solutions
using the \KBc\ algebra \cite{Okawa,ES}:
\begin{equation}
U=1-F(K)BcF(K) ,
\label{eq:U=1-FBcF}
\end{equation}
where $F(K)$ is a function of $K$, which should be carefully chosen to
realize a non-trivial solution. The inverse of $U$ is
\begin{equation}
U^{-1}=1+\frac{F}{1-F^2}\,BcF ,
\end{equation}
and the corresponding $\Psi$ is given by
\begin{equation}
\Psi=U\QB U^{-1}=FcK\frac{1}{1-F^2}BcF .
\label{eq:Psi=FcK...}
\end{equation}
For a given $F(K)$, we introduce an interpolating $F_s(K)$
which satisfies $F_{s\,=\,1}=F$ and $F_{s\,=\,0}=0$. Then,
$U_s$ and $\Psi_s$ are given by
\begin{equation}
U_s=1-F_s Bc F_s,
\qquad
\Psi_s=U_s\QB U_s^{-1}=F_scK\frac{1}{1-F_s^2}BcF_s .
\label{eq:U_s,Psi_s}
\end{equation}
For this $\Psi_s$, $\calB(s)$ of \eqref{eq:calB=another} is calculated
to give
\begin{align}
\calB(s)&=\int\deriv{}{s}\!\left(BcF_s^2c\frac{K}{1-F_s^2}\right)
\left(BcF_s^2c\frac{K}{1-F_s^2}\right)^2
=\sum_{a=1}^3\calB_a(s) ,
\label{eq:HcalB}
\end{align}
with
\begin{align}
\calB_1(s)&=\int\!BcK\!\left(\deriv{}{s}\frac{1}{1-F_s^2}\right)\!
\CR{c}{F_s^2}\frac{K}{1-F_s^2}cF_s^2c\frac{K}{1-F_s^2} ,
\nn\\
\calB_2(s)&=\int\!Bc\frac{K}{1-F_s^2}\CR{F_s^2}{c}\frac{K}{1-F_s^2}
c F_s^2 c\frac{K}{1-F_s^2}\deriv{F_s^2}{s} ,
\nn\\
\calB_3(s)&=\int\!BcK\!\left(\deriv{}{s}\frac{1}{1-F_s^2}\right)\!
\CR{F_s^2}{c}\frac{K}{1-F_s^2}c F_s^2 c\frac{KF_s^2}{1-F_s^2} .
\label{eq:HcalBa}
\end{align}
In this derivation, we have used the following identity due to the
\KBc\ algebra:
\begin{align}
\int\!Bc A_1 c A_2 Bc A_3 c A_4 Bc A_5 c A_6
&=\int\!Bc A_1 A_2 c A_3 A_4 c A_5 c A_6
-\int\!Bc A_1 A_2 A_3 c A_4 c A_5 c A_6
\nn\\
&
-\int\!B A_1 c A_2 c A_3 A_4 c A_5 c A_6
+\int\!B A_1 c A_2 A_3 c A_4 c A_5 c A_6 ,
\end{align}
where $A_k$ ($k=1,\cdots,6$) are arbitrary functions of $K$.
Of course, we get the same result if we use \eqref{eq:calB=intQB()}
for $\calB(s)$.
More elaborate manipulation using the \KBc\ algebra leads to
a simpler expression:
\begin{equation}
\calB(s)=\wt{\calB}_1(s)+\wt{\calB}_2(s) ,
\label{eq:KcalB}
\end{equation}
with
\begin{align}
\wt{\calB}_1(s)&=\int\!BcF_s^2 cK\left\{
c\frac{K}{1-F_s^2}\deriv{F_s^2}{s}c\frac{K}{1-F_s^2}
-\frac{1}{1-F_s^2}\deriv{F_s^2}{s}cK\CR{cK}{\frac{1}{1-F_s^2}}
\right\} ,
\nn\\
\wt{\calB}_2(s)&=-\int BcF_s^2cKc\frac{K}{\left(1-F_s^2\right)^2}
\deriv{F_s^2}{s}cK .
\label{eq:KcalBa}
\end{align}
$\wt{\calB}_1(s)$ and $\wt{\calB}_2(s)$ are separately given as
integrations of BRST-exact quantities.
Derivation of this expression is summarized in Appendix \ref{app:KcalB}.

\section{$\bm{\swN}$ for the tachyon vacuum}

In this section, we evaluate $\swN$ \eqref{eq:swN=intQBcalA} for $U$
representing the tachyon vacuum.
In particular, we show that, by using the \KeR\ mentioned in the
Introduction, $\swN$ reproduces the expected result $\swN=-1$.

\subsection{$\bm{\calB(s)}$ without regularization}

As $F(K)$ corresponding to the tachyon vacuum, we choose \cite{ES}
\begin{equation}
F^2=\frac{1}{1+K} ,
\label{eq:F^2_tv}
\end{equation}
and as the interpolating $F_s^2$ we take simply
\begin{equation}
F_s^2=s F^2=\frac{s}{1+K} .
\label{eq:F_s^2_tv}
\end{equation}
Let us consider calculating $\calB(s)$ given by \eqref{eq:HcalB}
or \eqref{eq:KcalB}. There appear in \eqref{eq:HcalBa} and
\eqref{eq:KcalBa} quantities $1/(1-s+K)^k$ ($k=1,2$) and $1/(1+K)$;
the former is from $1/(1-F_s^2)^k$, while the latter is $F^2$ itself.
For them we use the Schwinger parametrizations:
\begin{equation}
\frac{1}{(1-s+K)^k}=\frac{1}{(k-1)!}
\int_0^\infty\!dt\;t^{k-1}\,e^{-t(1-s+K)} .
\quad
\frac{1}{1+K}=\int_0^\infty\!d\wt{t}\,e^{-\wt{t}(1+K)},
\end{equation}
Then, we make a change of variables from $t_a$'s for $1/(1-s+K)^k$
and $\wt{t}_b$'s for $1/(1+K)$ to $(x,y,z_1,z_2,\cdots)$ satisfying
\begin{equation}
\sum_a t_a=x,
\quad
\sum_b\wt{t}_b=x\left(\frac{1}{y}-1\right),
\quad
\left(0\le x<\infty,\ 0\le y\le 1\right) .
\label{eq:sumt}
\end{equation}
The variables $(z_1,z_2,\cdots)$ are introduced for expressing $t_a$'s
and $\wt{t}_b$'s in such a way that they satisfy the constraints
\eqref{eq:sumt}.
An example which appears in the calculation of $\calB_1(s)$ in
\eqref{eq:HcalBa} is
\begin{align}
&\int\!Bc\frac{1}{(1-s+K)^2}\,c\,\frac{1}{1-s+K}\,c\,
\frac{1}{1+K}\,cK
\nn\\
&=\int_0^\infty dx\int_0^1\!dy\,e^{-x(1/y-s)}
\frac{x^2}{y^2}\int_0^1dz\,t_1
\left(-\Drv{}{t_4}\right)G(t_1,t_2,\wt{t}_3,t_4)
\biggr|_{t_1=xz,\,t_2=x(1-z),\,\wt{t}_3=x(1/y-1),\,t_4=0} ,
\label{eq:calcHcalB1}
\end{align}
where $x^2/y^2$ is the Jacobian of the change of variables
and $G(t_1,t_2,t_3,t_4)$ is given in terms of the correlator on the
cylinder with circumference $\sum_{a=1}^4 t_a$ by (see Appendix A)
\begin{equation}
G(t_1,t_2,t_3,t_4)=
\VEV{Bc(0)c(t_1)c(t_1+t_2)c(t_1+t_2+t_3)}_{t_1+t_2+t_3+t_4} .
\label{eq:G}
\end{equation}
Carrying out this kind of calculation for the whole of
\eqref{eq:HcalB} or \eqref{eq:KcalB}, we find that $\calB(s)$ is given
as an integration over $(x,y)$,
\begin{equation}
\calB(s)=\int_0^\infty\!dx\int_0^1\!dy\,e^{-x(1/y-s)}\,H(x,y) ,
\label{eq:calB=intdxdy}
\end{equation}
and moreover that $H(x,y)$ vanishes identically; $H(x,y)=0$.
This is consistent with the fact that $\calB(s)$
\eqref{eq:calB=intQB()} is the integration of a BRST-exact quantity.
Therefore, $\swN$ given by \eqref{eq:swN=intcalB} also vanishes.
On the other hand, we know that $\swN=-1$ from the direct calculation
of the energy density of the tachyon vacuum solution.
This contradiction will be resolved in the next subsection by
introducing the \KeR.

\subsection{$\bm{\calB(s)}$ in the $\bm{K_\eps}$-regularization}
\label{sec:calBinKeR}

As we explained in the Introduction, calculations of various
correlators suggest that the eigenvalue of the operator $K$ is
non-negative and, in particular, that there is a zero eigenvalue.
This is also seen from a concrete calculation of
\eqref{eq:calcHcalB1}; it is finite for $s<1$, while it diverges at
$s=1$ (and also for $s>1$).
Therefore, in order to make $\calB(s)$ non-vanishing and obtain
$\swN=-1$ from \eqref{eq:swN=intcalB}, it seems necessary to introduce a
regularization to $\calB(s)$ which extracts and at the same time
regularize the divergent contribution of the zero eigenvalue of $K$ to
$\calB(s)$. Without regularization, the zero eigenvalue would be
unseen due the BRST-exactness of the integrand of $\calB(s)$.
Such a regularization must fulfill two requirements:
First, it must regularize the divergence of each term in $\calB(s)$,
such as \eqref{eq:calcHcalB1}, at $s=1$ due to the
zero eigenvalue. Second, it must violate the BRST-exactness of the
integrand of $\calB(s)$ \eqref{eq:calB=intQB()}.
For example, introduction of the upper cutoff to the $x$-integration
in \eqref{eq:calcHcalB1} regularizes the divergence at $s=1$. However,
it does not violate the BRST-exactness, and $\calB(s)$ given by
\eqref{eq:calB=intdxdy} remains zero in this regularization since
$H(x,y)$ remains unchanged from zero.

As a regularization which can also violate the BRST-exactness of the
integrand of $\calB(s)$ \eqref{eq:calB=intQB()} and hence that of
$\swN$ \eqref{eq:swN=intQBcalA}, we adopt the \KeR\
\eqref{eq:KtoKe=K+eps} as we mentioned in the Introduction.
This is to replace all $K$'s in the integrand of $\calB(s)$
\eqref{eq:calB=intQB()} (and that of $\swN$ \eqref{eq:swN=intQBcalA})
with $\Ke=K+\eps$.
In particular, we must make the replacement $K\to\Ke$ {\em after}
calculating the operation of $\QB$.
Namely,
\begin{equation}
\calB(s)\Bigr|_\text{\Ker}
=\int\!\left[\QB\!\left(\Psi_s *\deriv{\Psi_s}{s}\right)
\right]_{K\to\Ke}
=\int\!\left(\deriv{\Psi_s}{s}*\Psi_s *\Psi_s\right)_{K\to\Ke} .
\end{equation}
For concrete calculations of the regularized $\calB(s)$, we use
\eqref{eq:HcalB} or \eqref{eq:KcalB} with all the $K$'s replaced with
$\Ke$.

Let us recalculate $\calB(s)$ for the tachyon vacuum in the \KeR\ by
using the expression \eqref{eq:HcalB}.
Note that each of $\calB_a(s)$ ($a=1,2,3$) \eqref{eq:HcalBa} is
given in the form
\begin{equation}
\int\!Bc W_1 c W_2 c W_3 c W_4 ,
\end{equation}
with $W_k$ ($k=1,2,3,4$) being functions of $K$ and $s$.
For the present $F_s^2$ \eqref{eq:F_s^2_tv}, $W_k$ is a rational
function with its denominator consisting of $1-s+K$ and $1+K$.
For each $W_k$ we take the expression where the numerators do not
contain any $K$. For example,
\begin{align}
\frac{K}{1-F_s^2}&=K+s-\frac{s(1-s)}{1-s+K} ,
\label{eq:K/(1-F_s^2)}
\\
K\left(\deriv{}{s}\frac{1}{1-F_s^2}\right)
&=1-\frac{1-2s}{1-s+K}-\frac{s(1-s)}{(1-s+K)^2} .
\end{align}
For later convenience, we call $K$ which is not contained in the
denominators (for example, the first $K$ on the RHS of
\eqref{eq:K/(1-F_s^2)}) ``bare $K$'' hereafter.
In the \KeR, all the $K$'s are replaced by $\Ke$.
However, a bare $K$ in $W_{1,2,3}$ which is sandwiched between two
$c$'s remains $K$ owing to $c^2=0$. For example, we have
\begin{equation}
\left.c\frac{K}{1-F_s^2}c\,\right|_{K\to\Ke}
=c\left(K+s-\frac{s(1-s)}{1-s+\Ke}\right)c .
\end{equation}
However, this is not the case for $W_4$ since it is not located
between two $c$'s.
Note that $W_4$ containing the bare $K$ is only $K/(1-F_s^2)$
\eqref{eq:K/(1-F_s^2)} which appears in $\calB_1$.
$W_4$ in $\calB_2$ and $\calB_3$ is given by
\begin{equation}
\frac{K}{1-F_s^2}\deriv{F_s^2}{s}
=\frac{1}{s}\frac{K F_s^2}{1-F_s^2}=1-\frac{1-s}{1-s+K} .
\end{equation}
Therefore, the regularized $\calB(s)$ consists of two parts
corresponding to the replacement of the bare $K$ in $W_4$ by
$\Ke=K+\eps$. One is the part from this $\eps$, and the other is all
the rest. We call the former proportional to $\eps$ ``$\eps$-term'',
and the latter ``non-$\eps$-term''. In both the terms, all the $K$'s
in the denominators are now replaced with $\Ke$.

First, we find that the non-$\eps$-term is equal to zero.
To see this, note that the effect of the replacement $K\to\Ke$ in a
denominator is to multiply its Schwinger parametrization by
$e^{-\eps t}$. Since the sum of all the Schwinger parameters is equal
to $x/y$ (see \eqref{eq:sumt}), we find that the whole of the
non-$\eps$-term is given simply by \eqref{eq:calB=intdxdy} with
$e^{-x(1/y-s)}$ replaced by $e^{-x\left[(1+\eps)/y-s\right]}$.
The non-$\eps$-term vanishes since the function $H(x,y)$ remains
unchanged from the $\eps=0$ case and is equal to zero.

Therefore, we have only to calculate the $\eps$-term.
As we explained above, only $\calB_1$ contributes to the $\eps$-term.
Denoting $\calB(s)$ in the \KeR\ by $\calB_\eps(s)$,
we have
\begin{align}
&\calB_\eps(s)
=\eps s^2\biggl\{
(1-s)^2\int\!c\frac{1}{1-s+\Ke}c\frac{1}{1-s+\Ke}c\frac{1}{1+\Ke}
\nn\\
&\qquad\qquad\qquad\qquad
-\int\!c\left[\frac{1}{1-s+\Ke}-\frac{1-s}{(1-s+\Ke)^2}\right]
cKc\frac{1}{1+\Ke}\biggr\}
\nn\\
&=\eps s^2\int_0^\infty\!dx\int_0^1\!dy\,
e^{-x\left[(1+\eps)/y-s\right]}
\biggl\{
(1-s)^2\,
\frac{x^2}{z^2}\int_0^1\!dz\,G_c(t_1,t_2,t_3)\Bigr|_{
t_1=xz,t_2=x(1-z),t_3=x(1/y-1)}
\nn\\
&\qquad
+\frac{x}{z^2}\left[1-(1-s)t_1\right]\Drv{}{t_2}
G_c(t_1,t_2,t_3)\Bigr|_{t_1=x,t_2=0,t_3=x(1/y-1)}
\biggr\}
\nn\\
&=\eps s^2\int_0^\infty\!dx\int_0^1\!dy\,
e^{-x\left[(1+\eps)/y-s\right]} f(x,y,s) ,
\label{eq:calccalB_tv}
\end{align}
where $G_c$ is the $ccc$ correlator on the cylinder (see
\eqref{eq:VEVccc}),
\begin{equation}
G_c(t_1,t_2,t_3)=\VEV{c(0)c(t_1)c(t_1+t_2)}_{t_1+t_2+t_3}
=G(t_1,t_2,t_3,t_4=0) ,
\end{equation}
and the function $f(x,y,s)$ is given by
\begin{equation}
f(x,y,s)=\frac{x^3\,\sin\pi y}{2\pi^4 y^6}
\Bigl\{\pi(1-s)^2 x^2 y\cos\pi y
-\bigl[(1-s)^2 x^2-2\pi^2(1-s)xy^2+2\pi^2y^2\bigr]\sin\pi y\Bigr\} .
\end{equation}
In the first expression of \eqref{eq:calccalB_tv}, we have
used \eqref{eq:KBcA} to eliminate $B$.

It seems difficult to carry out explicitly the $(x,y)$ integrations in
\eqref{eq:calccalB_tv} to obtain an analytic expression of
$\calB_\eps(s)$ for a finite $\eps$. However, we can exactly evaluate
$\swN$ \eqref{eq:swN=intcalB} by carrying out first the $s$
integration:
\begin{equation}
\swN=\lim_{\eps\to 0}\pi^2\int_0^1\!ds\,\calB_\eps(s)
=-\lim_{\eps\to 0}\frac{1}{(1+\eps)^3}=-1 .
\label{eq:swN=-1}
\end{equation}
This is the desired result for the tachyon vacuum.
Next, let us consider $\calB_\eps(s)$ itself given by
\eqref{eq:calccalB_tv}.
First, we see that $\lim_{\eps\to 0}\calB_\eps(s)=0$ for
$s<1$. This is understood from the facts that \eqref{eq:calccalB_tv}
is multiplied by $\eps$, and that the denominator $1-s+\Ke$
is positive definite for $s<1$ even when $\eps=0$.
In order to obtain the expression of $\calB_\eps(s)$ near $s=1$ for an
infinitesimal $\eps$, we make a change of variables from $(x,y)$ to
$(\xi,\eta)$ defined by
\begin{equation}
x=\frac{\xi}{\eps},\qquad
y=\left(1+\frac{\eps}{1+\eps}\eta\right)^{-1} .
\end{equation}
In addition, since we are interested in $s\simeq 1$, we use, instead
of $s$, the variable $w$:
\begin{equation}
s=1-\eps w .
\end{equation}
In terms of the new variables, the exponential function in the last
expression of \eqref{eq:calccalB_tv} is simply given by
$e^{-\xi(\eta+w+1)}$.
Then, we obtain
\begin{align}
\calB_\eps(s)&=\frac{\eps s^2}{1+\eps}
\int_0^\infty\!d\xi\int_0^\infty\!d\eta
\,e^{-\xi(\eta+w+1)}\,y^2f(x,y,s)
\nn\\
&=-\frac{3s^2}{\pi^2}\frac{w^2}{\eps(1+w)^4}
=-\frac{s^2}{\pi^2}\frac{3\eps(1-s)^2}{(1-s+\eps)^4}
\underset{\eps\to +0}{\to} -\frac{1}{\pi^2}\delta(1-s),
\label{eq:calB_eps(s)=}
\end{align}
where we have used
\begin{equation}
y^2 f(x,y,s)=-\frac{\eta w^2\xi^5}{2\pi^2\eps^2}+O(1/\eps) .
\end{equation}
In obtaining the final expression of \eqref{eq:calB_eps(s)=}, we used
that the $\eps\to +0$ limit of
\begin{equation}
\delta_\eps(1-s)=\frac{3\eps(1-s)^2}{(1-s+\eps)^4} ,
\end{equation}
can be identified as a delta function $\delta(1-s)$ in the interval
$0\le s\le 1$ since it satisfies
\begin{equation}
\lim_{\eps\to +0}\delta_\eps(1-s)=0\quad (s<1),
\qquad
\lim_{\eps\to +0}\int_0^1\!ds\,\delta_\eps(1-s)=1 .
\end{equation}
Our result \eqref{eq:calB_eps(s)=} implies that $\swN$
\eqref{eq:swN=intcalB} given as the integration of $\calB(s)$ has a
contribution only at $s=1$.
This reconfirms our earlier expectation that it is the zero eigenvalue of
$K$ that makes $\swN$ non-trivial.

\section{Additivity of $\bm{\swN}$ and the EOM}

As stated in the Introduction, we can easily derive the following
identity for $\swN$ \eqref{eq:swN}:
\begin{equation}
\swN[UV]=\swN[U]+\swN[V]+\Delta\swN ,
\label{eq:N12=N1+N2+DN}
\end{equation}
with $\Delta\swN$ given by
\begin{equation}
\Delta\swN=\pi^2\int\!\QB\Bigl\{
\left(\QB U^{-1}\right)UV\left(\QB V^{-1}\right)\Bigr\} .
\label{eq:DN}
\end{equation}
The same kind of equations as
\eqref{eq:N12=N1+N2+DN} and \eqref{eq:DN} hold for the
winding number \eqref{eq:wN} in the CS theory.
Eq.\ \eqref{eq:N12=N1+N2+DN} leads to the additive law of $\swN$,
\begin{equation}
\swN[UV]=\swN[U]+\swN[V] ,
\label{eq:N12=N1+N2}
\end{equation}
if we can discard the last term $\Delta\swN$ which is the integration
of a BRST-exact quantity.
The additivity \eqref{eq:N12=N1+N2} means, in particular, that
$(\Utv)^{n}$ ($n=\pm 1,\pm 2,\ldots$) with $\Utv$ of the tachyon
vacuum has an integer $\swN$:
\begin{equation}
\swN[(\Utv)^n]=n\,\swN[\Utv]=-n .
\label{eq:swN[U_n]=-n}
\end{equation}
In the above argument, we did not take the regularization into
account. In this section, we will examine whether
\eqref{eq:swN[U_n]=-n} really holds in our \KeR.

\subsection{Calculation of $\bm{\swN[(\Utv)^n]}$}
\label{sec:U_n}

In this subsection, we calculate $\swN$ for $U=(\Utv)^n$ in the \KeR.
For this purpose, we first obtain $F^2$ corresponding to $(\Utv)^n$.
Let us start with a generic $U$ given in the form
\eqref{eq:U=1-FBcF}. Then, using the \KBc\ algebra, we find that
\begin{equation}
U^n=1-\wt{F}_n BcF,\qquad (n=0,\pm 1,\pm 2,\cdots),
\label{eq:U^n=1-F_nBcF}
\end{equation}
with $\wt{F}_n$ being a function of $K$ only
($\wt{F}_0=0$, $\wt{F}_1=F$), and further that $\wt{F}_n$ satisfies
the following recursion relation:
\begin{equation}
\wt{F}_{n+1}=F+\wt{F}_n(1-F^2) .
\label{eq:RRwtF_n}
\end{equation}
Though $U^n$ given by \eqref{eq:U^n=1-F_nBcF} is not of the standard
form \eqref{eq:U=1-FBcF}, we can bring it into the standard form in
terms of $R_n$ which is a function of $K$ only:
\begin{equation}
R_n U^n R_n^{-1}=1-F_n Bc F_n .
\label{eq:U_n}
\end{equation}
Note that $R_nU^n R_n^{-1}$ carries the same $\swN$ as $U^n$ since
$R_n$ commutes with $\QB$. From \eqref{eq:U^n=1-F_nBcF} and
\eqref{eq:U_n}, we have $R_n\wt{F}_n=F_n=F R_n^{-1}$ and hence
$F_n^2=F\wt{F}_n$.
Then, from the recursion relation \eqref{eq:RRwtF_n}, we obtain
$1-F_{n+1}^2=(1-F^2)(1-F_n^2)$  and therefore
\begin{equation}
F_n^2=1-(1-F^2)^n ,\qquad (n=0,\pm 1,\pm 2,\cdots).
\label{eq:F_n^2=}
\end{equation}

In particular, by taking as $F^2$ that for the tachyon vacuum,
\eqref{eq:F^2_tv}, we obtain
\begin{equation}
F_n^2=1-\left(\frac{K}{1+K}\right)^n .
\label{eq:F_n^2ofMS}
\end{equation}
This coincides with $F^2$ proposed in \cite{MS} as an example giving
$\swN=-n$ from quite a different argument.

We have calculated $\swN[(\Utv)^n]$ by using $F_n^2$ given by
\eqref{eq:F_n^2ofMS} in the \KeR.
As the interpolating $(F_n^2)_s$ with parameter $s$, we adopt
\eqref{eq:F_n^2=} with $F^2$ replaced by the interpolating $F^2_s$
\eqref{eq:F_s^2_tv} for the tachyon vacuum:
\begin{equation}
(F^2_n)_s=1-\left(\frac{1-s+K}{1+K}\right)^n .
\end{equation}
The calculations are almost the same as in the case of the tachyon
vacuum except that, for a negative $n$, there also appear terms which
contain $1/(1-s+K)$ but no $1/(1+K)$.
For such terms the variable $y$ in \eqref{eq:sumt} is unnecessary, and
they are reduced to a single integration of the form
\begin{equation}
\int_0^\infty\!dx\,e^{-(1-s)x} J(x) .
\label{eq:intJ}
\end{equation}
Now the total of $\calB(s)$ is given as the sum of two types of
integrations, \eqref{eq:calB=intdxdy} and \eqref{eq:intJ}.
The point is that each of $H(x,y)$ and $J(x)$ separately vanishes
before introducing the regularization since they separately come from
the integration of a BRST-exact quantity. Therefore, in obtaining
$\calB(s)$ in the \KeR, we are allowed to take only the $\eps$-terms
as in the case of the tachyon vacuum explained in Sec.\
\ref{sec:calBinKeR}.

Our results for $n=\pm 1$ and $\pm 2$ are already given in
\eqref{eq:swN[U_n]=}.
It shows that our expectation \eqref{eq:swN[U_n]=-n} does not hold
except in the cases $n=\pm 1$. Even worse, $\swN$ for $n=\pm 2$ are
not integers.
This result \eqref{eq:swN[U_n]=} implies that the additivity
\eqref{eq:N12=N1+N2} is violated. We have examined the extra term
$\Delta\swN$ \eqref{eq:DN} in the \KeR, namely,
$\Delta\swN|_\text{\Ker}
=\pi^2\int\left(\QB\left\{\cdots\right\}\right)_{K\to\Ke}$
with $U=V=(\Utv)^{\pm 1}$ to confirm that it is non-vanishing and
exactly accounts for the violation of the additivity
$\swN[(\Utv)^{\pm 2}]=2\,\swN[(\Utv)^{\pm1}]$.

On the other hand,
\begin{equation}
\swN[U^{-1}]=-\swN[U] ,
\label{eq:swN[U^-1]=-swN[U]}
\end{equation}
holds for any $U$, since we have
$\Delta\swN|_\text{\Ker}
=\pi^2\int\!\left(\QB^2(U^{-1}\QB U)\right)_{K\to\Ke}=0$ for
\eqref{eq:N12=N1+N2+DN} and \eqref{eq:DN} with $UV=1$ owing to
$\QB^2=0$. Our result \eqref{eq:swN[U_n]=} is consistent with the
property \eqref{eq:swN[U^-1]=-swN[U]}.

\subsection{EOM in the strong sense}

Let us check whether $\Psi=U\QB U^{-1}$ with $U=(\Utv)^n$ satisfies
the EOM in the strong sense,
$\int\!\Psi*\left(\QB\Psi+\Psi^2\right)=0$, in the \KeR.
Since this EOM is the same between $U=(\Utv)^n$ and
$R_n(\Utv)^nR_n^{-1}$, we consider the latter with $F_n^2$ given by
\eqref{eq:F_n^2ofMS}.

For this purpose, we prepare the expression of the EOM for a generic
$U$ in the standard form \eqref{eq:U=1-FBcF}.
Let $\Psie$ be the $\Ke$-regularized $\Psi$,
\begin{equation}
\Psie=\left(U\QB U^{-1}\right)_{K\to\Ke}
=\Fe c\frac{\Ke B}{1-\Fe^2}c\Fe ,
\label{eq:Psie}
\end{equation}
with $\Fe=F(\Ke)$. Using the \KBc\ algebra, we find that the EOM of
$\Psie$ is reduced to an apparently of $O(\eps)$ quantity:
\begin{equation}
\QB\Psie+\Psie *\Psie=\eps\times\Fe c\frac{\Ke}{1-\Fe^2}c\Fe ,
\end{equation}
where $\QB\Psie$ is $\Psie$ \eqref{eq:Psie} acted by $\QB$, and
is not equal to $(\QB\Psi)_{K\to\Ke}$.
Using this for $F=F_n$ \eqref{eq:F_n^2ofMS}, a straightforward
calculation gives
\begin{equation}
\int\!\Psie *\left(\QB\Psie+\Psie *\Psie\right)
=\eps\int\!Bc\Fe^2c\frac{\Ke}{1-\Fe^2}
c\Fe^2c\frac{\Ke}{1-\Fe^2}
\underset{\eps\to 0}{\to}
\begin{cases}
-6 &(n=-2)\\ 0 & (n=\pm 1) \\ 2 &(n=2)
\end{cases} .
\label{eq:EOMtest}
\end{equation}

Our result that $\Psie$ for $n=\pm 2$ does not satisfy the EOM in the
strong sense implies that this $\Psie$ cannot be regarded as a
pure-gauge even in the limit $\eps\to 0$.
$\swN$ \eqref{eq:swN} is formally invariant under small
deformations of $U$ owing to the fact that $\Psi=U\QB U^{-1}$ is a
pure-gauge.
The violation of the EOM for $n=\pm2$, \eqref{eq:EOMtest}, means that
$\swN[(\Utv)^{\pm 2}]$ is not such a stable quantity.
Therefore, the anomalous (non-integer) values of $\swN$ presented in
\eqref{eq:swN[U_n]=}, $\swN[(\Utv)^{\pm 2}]=\mp(2-2\pi^2)$,
should not be taken as a counterexample to the quantization of $\swN$.

This hand-waving argument should, of course, be made more rigorous.
In particular, we must clarify the relationships among the various
requirements: the EOM, the additive law \eqref{eq:N12=N1+N2} of
$\swN$, inertness of $\swN$ under deformations of $U$, and the
quantization of $\swN$.
Here, we examined the validity of the EOM,
\begin{equation}
\int\calO*\left(\QB\Psie+\Psie *\Psie\right)=0 ,
\end{equation}
only for $\calO=\Psie$. It is necessary to understand for what class
of $\calO$ the EOM should hold in order for the requirements on $\swN$
to be valid.

Finally, a comment is in order concerning a simpler derivation of
\eqref{eq:EOMtest}. In the above, it was evaluated without any
approximation. However, the same result can be obtained by taking only
the term with the least power of $\Ke$ in the Laurent series of each
of the quantities $\Fe^2$ and $\Ke/(1-\Fe^2)$ with respect to $\Ke$.
For example, for $n=2$, we have
\begin{equation}
\Fe^2=1-\Ke^2+O(\Ke^3),\qquad
\frac{\Ke}{1-\Fe^2}=\frac{1}{\Ke}+O(\Ke^0) ,
\end{equation}
and \eqref{eq:EOMtest} for $n=2$ is reproduced by
\begin{equation}
\eps\int\!Bc\Ke^2 c\frac{1}{\Ke}c\Ke^2 c\frac{1}{\Ke}=2 .
\end{equation}

\section{Summary and discussions}

In this paper, motivated by the similarity between the CSFT and the CS
theories, we pursued the possibility that $\swN$ \eqref{eq:swN} is
interpreted as a kind of winding number in CSFT which is quantized to
integer values. We especially focused on the expression
\eqref{eq:swN=intQBcalA} of $\swN$ as the integration of a BRST-exact
quantity, which naively vanishes identically and manifests the
topological nature of $\swN$. For realizing non-vanishing values of
\eqref{eq:swN=intQBcalA}, we need to introduce a regularization for
divergences arising from the zero eigenvalue of the operator $K$.
This regularization must also cause an infinitesimal violation of the
BRST-exactness of the integrand of $\swN$.
As such a regularization, we proposed the \KeR\ \eqref{eq:KtoKe=K+eps}
of shifting $K$ by a positive infinitesimal $\eps$.
Applying the \KeR\ to the calculation of \eqref{eq:swN=intQBcalA} for
$U=\Utv$ which represents the tachyon vacuum, we got the expected
result $\swN[\Utv]=-1$. In this calculation, we found that the
non-vanishing value of $\swN$ is realized by $\eps\times(1/\eps)$ with
$\eps$ from the violation of the BRST-exactness of the integrand and
$1/\eps$ from the zero eigenvalue of $K$.
Then, we further studied $\swN$ for $U=(\Utv)^n$  with $n=-2,-1,2$.
The additive law of $\swN[U]$ for the product of $U$ predicts that
$\swN[(\Utv)^n]=-n$. However, explicit calculations show that $\swN$
for $n=\pm 2$ are anomalous as given in \eqref{eq:swN[U_n]=}.
At the same time, we found that $\Psi=U\QB U^{-1}$ with
$U=(\Utv)^{\pm 2}$ does not satisfy the EOM in the strong sense
either. This implies that $U\QB U^{-1}$ for such $U$ cannot be
regarded as truly pure-gauge, and may explain the violation of the
quantization of $\swN$.

This paper is a first step toward identifying $\swN$ as a winding
number in CSFT and thereby unveiling the ``topological structure'' of
CSFT. Our analysis is of course far from being complete and there
remains many open questions to be answered.
They include the followings:

\begin{itemize}
\item
We attributed our unwelcome result that $\swN[U=(\Utv)^{\pm 2}]$ take
non-integer values to the breakdown of the EOM in the strong sense
for $\Psi=U\QB U^{-1}$. However, we do not know a precise connection
between the two. We have to understand the relationships among the
quantization of $\swN$, invariance of $\swN$ under small deformations
of $U$, the additive law of $\swN$, and the EOM in the strong sense.

\item
In this paper, we proposed and used the \KeR.
This regularization certainly regularizes the infinities arising from
the zero eigenvalue of $K$ and, at the same time, violates the
BRST-exactness of the integrand of $\swN$ \eqref{eq:swN=intQBcalA},
thus leading to the desired value $\swN[\Utv]=-1$.
However, we do not know whether our \KeR\ is a fully satisfactory
one. It might be that the non-integer values of
$\swN[U]$ and the violation of EOM for $U=(\Utv)^{\pm 2}$ are
artifacts of the \KeR. We need to understand the basic principles that
the regularization has to satisfy.

\item
Besides such general considerations as presented above, it is an
interesting problem to construct $U$'s which give integer $\swN$ other
than $\pm 1$, and at the same time, satisfy the EOM in the strong
sense.

\item
The existence of $U$ with $\swN[U]<-1$ apparently implies a physically
unwelcome fact that $\Psi=U\QB U^{-1}$ represents a state with its
energy density lower than that of the tachyon vacuum.
We have to show that, if there exists such $U$, $\Psi=U\QB U^{-1}$
never satisfies the EOM in the strong sense.

\item
In this paper, we calculated $\swN[(\Utv)^n]$ only for $n=\pm 1,\pm2$.
It is an interesting technical problem to obtain its expression for a
generic integer $n$.

\item
It is a challenging problem to evaluate $\swN$ as a ``surface
integration'', $\swN=\int_\calM\QB\calA=\int_{\p\calM}\calA$.
For this we have to understand the meaning of the boundary $\p\calM$
(which should be a set of singularities of $\calA$) as well as that of
the original ``manifold'' $\calM$.
We have to understand of course what kind of ``windings'' the quantity
$\swN$ counts.
\end{itemize}
By resolving these problems, we wish to find fruitful structure of
SFT which we still do not know.

\section*{Acknowledgments}
We would like to thank Isao Kishimoto, Toru Masuda, Masaki Murata,
Toshifumi Noumi, Tomohiko Takahashi and Barton Zwiebach for valuable
discussions.
T.\ K.\ thanks Maiko Kohriki, Taichiro Kugo and Hiroshi Kunitomo for
teaching her the basics of SFT.
The work of H.~H.\ was supported in part by a Grant-in-Aid for
Scientific Research (C) No.~21540264 from the Japan Society for the
Promotion of Science (JSPS).

\vspace{5mm}
\noindent
{\large\bf Note added}\\
The \KeR\ was also used in \cite{EM} and \cite{MSSFT2011} in their
analysis of the energy and the EOM of classical solutions.
We would like to thank Ted Erler and Masaki Murata for
correspondences.

\newpage
\appendix

\section{$\bm{KBc}$ algebra and correlators}
\label{app:KBc}

Here, we summarize the \KBc\ algebra and the correlators which are
used in the text. (See \cite{Okawa,ES} for details. In this
paper, we follow the convention of \cite{ES}.)

The elements of the \KBc\ algebra satisfy
\begin{equation}
\CR{B}{K}=0,\quad\left\{B,c\right\}=1,\quad
B^2=c^2=0,
\label{eq:KBcA}
\end{equation}
and
\begin{equation}
\QB B=K,\quad\QB K=0,\quad\QB c=cKc .
\end{equation}
Their ghost numbers are
\begin{equation}
g(K)=0,\quad g(B)=-1,\quad g(c)=1 .
\end{equation}

In the text, there appeared the following quantities:
\begin{align}
G(t_1,t_2,t_3,t_4)
&=\int\!Bc\,e^{-t_1 K}c\,e^{-t_2 K}c\, e^{-t_3 K}c\,e^{-t_4 K}
\nn\\
&=\VEV{Bc(0)c(t_1)c(t_1+t_2)c(t_1+t_2+t_3)}_{t_1+t_2+t_3+t_4},
\\[3mm]
G_c(t_1,t_2,t_3)&=\int\!c\,e^{-t_1 K}c\,e^{-t_2 K}c\,e^{-t_3 K}
=\VEV{c(0)c(t_1)c(t_1+t_2)}_{t_1+t_2+t_3}.
\end{align}
They are given in terms of the correlators on the cylinder with
infinite length and the circumference $\ell$:
\begin{align}
\VEV{B\,c(z_1)c(z_2)c(z_3)c(z_4)}_{\ell}
&=\left(\frac{\ell}{\pi}\right)^2\biggl\{
-\frac{z_1}{\pi}\sin\!\left[\frac{\pi}{\ell}(z_2-z_3)\right]
\sin\!\left[\frac{\pi}{\ell}(z_2-z_4)\right]
\sin\!\left[\frac{\pi}{\ell}(z_3-z_4)\right]
\nn\\
&\qquad
+\frac{z_2}{\pi}\sin\!\left[\frac{\pi}{\ell}(z_1-z_3)\right]
\sin\!\left[\frac{\pi}{\ell}(z_1-z_4)\right]
\sin\!\left[\frac{\pi}{\ell}(z_3-z_4)\right]
\nn\\
&\qquad
-\frac{z_3}{\pi}\sin\!\left[\frac{\pi}{\ell}(z_1-z_2)\right]
\sin\!\left[\frac{\pi}{\ell}(z_1-z_4)\right]
\sin\!\left[\frac{\pi}{\ell}(z_2-z_4)\right]
\nn\\
&\qquad
+\frac{z_4}{\pi}\sin\!\left[\frac{\pi}{\ell}(z_1-z_2)\right]
\sin\!\left[\frac{\pi}{\ell}(z_1-z_3)\right]
\sin\!\left[\frac{\pi}{\ell}(z_2-z_3)\right]\biggr\} ,
\label{eq:VEVBcccc}
\\[3mm]
\VEV{c(z_1)c(z_2)c(z_3)}_{\ell}&=\left(\frac{\ell}{\pi}\right)^3
\sin\left[\frac{\pi}{\ell}(z_1-z_2)\right]
\sin\left[\frac{\pi}{\ell}(z_1-z_3)\right]
\sin\left[\frac{\pi}{\ell}(z_2-z_3)\right] .
\label{eq:VEVccc}
\end{align}

\section{Eq.\ (\ref{eq:KcalB}) and rederivation of  (\ref{eq:swN=-1})}
\label{app:KcalB}

In this appendix, we outline the derivation of another and simpler
expression \eqref{eq:KcalB} for $\calB(s)$ and the calculation of
$\swN$ for the tachyon vacuum using this expression.

First, $\calB(s)$ is given in the following form:
\begin{align}
\calB(s)&=\int\!BcF_s^2c\frac{K}{1-F_s^2}
\left(\deriv{}{s}BcF_s^2c\frac{K}{1-F_s^2}\right)
BcF_s^2c\frac{K}{1-F_s^2}
\nn\\
&=-\int\!Bc\left[cK,F_s^2\right]\frac{1}{1-F_s^2}
\left(\deriv{}{s}Bc\CR{cK}{F_s^2}\frac{1}{1-F_s^2}\right)
Bc\left[cK,F_s^2\right]\frac{1}{1-F_s^2}
\nn\\
&=-\int\!Bc\left[cK,F_s^2\right]\frac{1}{1-F_s^2}
\left(\deriv{}{s}\CR{cK}{F_s^2}\frac{1}{1-F_s^2}\right)
\left[cK,F_s^2\right]\frac{1}{1-F_s^2} ,
\end{align}
where we have used $c^2=0$ at the second equality.
The last expression is due to
\begin{equation}
\int\!Bc\CR{cK}{A_1}A_2 Bc\CR{cK}{A_3}A_4 Bc\CR{cK}{A_5}A_6
=\int\!Bc\CR{cK}{A_1}A_2\CR{cK}{A_3}A_4 \CR{cK}{A_5}A_6 ,
\end{equation}
which is valid for arbitrary $A_k$'s depending only on $K$.
Then, using
$(1-F_s^2)^{-1}\CR{cK}{F_s^2}(1-F_s^2)^{-1}=\CR{cK}{(1-F_s^2)^{-1}}$,
$\calB(s)$ is further rewritten as follows:
\begin{align}
\calB&=\int BcF_s^2cK\left\{
\frac{1}{1-F_s^2}\left[cK,\deriv{F_s^2}{s}\right]
+\left[cK,\frac{1}{1-F_s^2}\right]\deriv{F_s^2}{s}\right\}
\left[cK,\frac{1}{1-F_s^2}\right]
\nn\\
&=\int BcF_s^2cK\CR{cK}{\frac{1}{1-F_s^2}\deriv{F_s^2}{s}}
\left[cK,\frac{1}{1-F_s^2}\right] .
\end{align}
Expanding the commutators, we get \eqref{eq:KcalB}.
Each of $\calB_1(s)$ and $\calB_2(s)$ is given as the integration of a
BRST-exact quantity:
\begin{align}
\wt{\calB}_1(s)&=\int\!\QB\!\left[
Bc F_s^2 c\frac{K}{1-F_s^2}\deriv{F_s^2}{s}c\frac{K}{1-F_s^2}\right],
\\
\wt{\calB}_2(s)&=-\int\!\QB\left[
Bc F_s^2 c\frac{K}{(1-F_s^2)^2}\deriv{F_s^2}{s}cK\right].
\end{align}
{}From this it is manifest that $\calB(s)$ without regularization
vanishes.

Then, let us consider evaluating $\calB(s)$ using the expression
\eqref{eq:KcalB} in the \KeR. As explained in Sec.\
\ref{sec:calBinKeR}, we have only to take the $\eps$-term.
Differently to the case of \eqref{eq:HcalB}, all the terms in
\eqref{eq:KcalB} contribute to the $\eps$-term. We get
\begin{align}
\calB_\eps(s)&=\eps s^2\left\{
(1-s)^2\int c\frac{1}{1+\Ke}c\frac{1}{1-s+\Ke}c\frac{1}{1-s+\Ke}
\right.
\nn\\
&\left.\quad\quad\quad\quad
-\int c\frac{1}{1+\Ke}cKc\left[
\frac{1}{1-s+\Ke}-\frac{1-s}{(1-s+\Ke)^2}\right]\right\} .
\end{align}
This leads to exactly the same result as the final expression of
\eqref{eq:calccalB_tv}.

\end{document}